\documentclass[twocolumn,prb,superscriptaddress]{revtex4}

\usepackage{graphicx}% Include figure files
\usepackage{dcolumn}% Align table columns on decimal point
\usepackage{bm}% bold math
\usepackage{latexsym}
\usepackage{color}
\usepackage{hyperref}
\usepackage{amsmath}
\usepackage{amssymb}
\usepackage{natbib}

\begin{document}

\title[APL]{Superconducting nano-mechanical diamond resonators} %Title of paper

\author{Tobias Bautze}
\email[]{tobias.bautze@grenoble.cnrs.fr} \affiliation{Univ.
Grenoble Alpes, Inst. NEEL, F-38042 Grenoble, France}
\affiliation{CNRS, Inst. NEEL, F-38042 Grenoble, France}
\author{Soumen Mandal}
\altaffiliation{Now at: School of Physics and Astronomy, Cardiff
University, Cardiff} \affiliation{Univ. Grenoble Alpes, Inst.
NEEL, F-38042 Grenoble, France} \affiliation{CNRS, Inst. NEEL,
F-38042 Grenoble, France}
\author{Oliver A. Williams}
\affiliation{Fraunhofer-Institut f\"ur Angewandte Festk\"{o}rperphysik, Tullastra{\ss}e 72, 79108 Freiburg, Germany}
\affiliation{University of Cardiff, School of Physics and Astronomy, Queens Buildings, The Parade, Cardiff CF24 3AA, United Kingdom}
\author{Pierre Rodi\`ere}
\affiliation{Univ. Grenoble Alpes, Inst. NEEL, F-38042 Grenoble,
France} \affiliation{CNRS, Inst. NEEL, F-38042 Grenoble, France}
\author{Tristan Meunier}
\affiliation{Univ. Grenoble Alpes, Inst. NEEL, F-38042 Grenoble,
France} \affiliation{CNRS, Inst. NEEL, F-38042 Grenoble, France}
\author{Christopher B\"{a}uerle}
\email[]{bauerle@grenoble.cnrs.fr} \affiliation{Univ. Grenoble
Alpes, Inst. NEEL, F-38042 Grenoble, France} \affiliation{CNRS,
Inst. NEEL, F-38042 Grenoble, France}

\date{\today}

%%%%%%%%%%%%%%%%%%%%

\begin{abstract}
In this work we present the fabrication and characterization of
superconducting nano-mechanical resonators made from
nanocrystalline boron doped diamond (BDD). The oscillators can be
driven and read out in their superconducting state and show
quality factors as high as 40,000 at a resonance frequency of
around 10 MHz. Mechanical damping is studied for magnetic fields
up to 3 T where the resonators still show superconducting
properties. Due to their simple fabrication procedure, the devices
can easily be coupled to other superconducting circuits and their
performance is comparable with state-of-the-art
technology.\end{abstract}

%%%%%%%%%%%%%%%%%%%%%

\keywords{NEMS, superconductivity, diamond, BDD, quality factor, superconducting circuits}
%Use showkeys class option if keyword
%display desired

\maketitle

\section{\label{sec:intro}Introduction}

%%%%

Nano-mechanical resonators allow to explore a variety of physical
phenomena. From a technological point of view, they can be used
for ultra-sensitive mass \cite{Yang2006, Jensen2008, Chaste2012},
force\cite{Braginskii1977, Ekinci2005, Moser2013}, charge
\cite{Steele2009, Lassagne2009} and displacement detection. On the
more fundamental side, they offer fascinating perspectives for
studying macroscopic quantum systems. Significant progress has
been made in the last few years by cooling a nano-mechanical
resonator into its ground state \cite{OConnell2010, Rocheleau2010,
Teufel2011, Chan2011}. Couplings between nano-mechanical
resonators and superconducting circuits have been realized and
even the creation of entanglement with these macroscopic
oscillators seems in reach \cite{Walter2013}. In order to exploit
such a system in quantum information technology, nano-mechanical
systems will have to be coupled to other quantum systems such as
light \cite{Safavi-Naeini2013} or superconducting circuits
\cite{OConnell2010, Armour2002, Naik2006, Regal2008, LaHaye2009,
Suh2010} and new materials to improve the coupling for such hybrid
systems are of importance. In this respect diamond is an extremely
attractive material. Despite the fact that diamond has exceptional
mechanical properties \cite{Wang2004, Gaidarzhy2007,
Ovartchaiyapong2012, Najar2013}, it has a relatively high
refractive index which allows to couple it to light. In addition
when doped with boron, it can be rendered superconducting with
remarkable electrical properties and makes it a promising material
for fully integrated hybrid nano-mechanical systems.

In this article we present the fabrication and characterization of
nano-mechanical resonators built out of superconducting diamond.
We demonstrate a simple top down process to fabricate these
resonators by common electron beam lithography and hence offer a
simple way to be implemented into superconducting circuits. We
compare their performance to state-of-the-art resonators and
investigate the limits of their superconductivity. Quality factors
around 40000 at around 10 MHz resonance frequency are demonstrated
and it is shown that it is possible to directly read out a
superconducting resonator at a magnetic field as high as 3 Tesla.

%%%%%%%%%%%%%%%%%%%%%%%%%%%%%%%%
\section{\label{sec:fab}Fabrication}
%%%%%%%%%%%%%%%%%%%%%%%%%%%%%%%%

The nano-mechanical resonators have been fabricated from a
superconducting nanocrystaline diamond film, grown on a silicon
wafer with a 500 nm thick SiO$_2$ layer. To be able to grow
diamond on the Si/SiO$_2$ surface, small diamond particles of a
diameter smaller than 6 nm are seeded onto the silica substrate
with the highest possible density \cite{Williams2011}. A
subsequent microwave plasma chemical-vapor deposition (CVD) allows
to grow and control various properties of the film. It is possible
to vary the grain sizes from few nanometers to few microns by
controlling the methane concentration. Furthermore one can add a
variety of dopants to drastically change some of the key
properties of pure diamond. While adding boron gas during the CVD
process one can turn the elsewise insulating diamond film metallic
and above a critical concentration even superconducting
\cite{Ekimov2004, Bustarret2004}. A detailed description of this
growth process can be found in reference \cite{Williams2011}.

%%%%%%%%%%%%%%%%%%%%%%%%%%%%%%%%%%%%%
%Fig. 1:beam %%%%%%%%%%%%%%%%%%%%%%%%%%%%%%%%%%%%%
\begin{figure}[htbp]
\centerline{\includegraphics[width=8cm,angle=0]{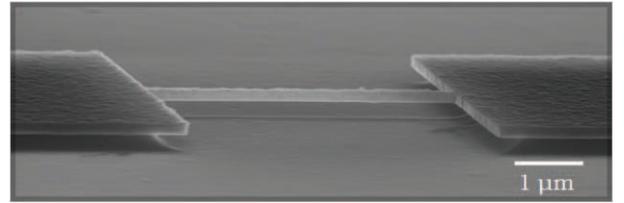}}
\caption{Scanning electron micro-graph of a diamond resonator. The
dimensions of the resonator are 480 nm x 300 nm (width x
thickness). Different resonators with different lengths ranging
from 5 to 30 $\mu$m have been fabricated.} \label{1sem}
\end{figure}
%%%%%%%%%%%%%%%%%%%%%%%%%%%%%%%%%%%%%%%%%%%%%%%

The nano-mechanical structures have been defined using standard
electron beam lithography. First, a 70 nm thick nickel etch mask
which defines the sample geometry has been evaporated on top of
the diamond film which has then been followed by anisotropic
oxygen plasma etching \cite{Mandal2011}. The sample is cooled to
10$^o$C while being exposed to the oxygen plasma for approximately
8 minutes. The anisotropy of the etching process leads to straight
walls that broaden less than 5 nm after 300 nm of etching. After
this process, the nickel mask is removed by dipping the sample in
a FeCl$_3$ solution. To provide good ohmic contacts, a tri-layer
consisting of titanium, platinum and gold has been evaporated
followed by annealing at 750$^o$C. The structures have been
suspended by etching the sacrificial SiO$_2$ layer using HF vapor
at 50$^o$C and atmospheric pressure for about 10 minutes. Diamond
itself is inert to this chemical etching process. Due to the
strong mechanical stiffness, tri-critical point drying has not
been necessary even for structures with lengths up to 30
micrometers. Figure \ref{1sem} displays a suspended diamond
structure that has been fabricated by this method. We have also
fabricated superconducting diamond resonators that were covered
with a 50 nm thick gold layer in order to test the samples at
temperatures above the superconducting transition temperature and
at currents above the superconducting critical current. This
allows for easy detection of the resonance conditions.  For this
purpose a slightly modified technique was used. After the e-beam
process a metallic bilayer of 50 nm gold and 50 nm of nickel was
deposited instead of 70 nm nickel as in the previous case. This
bilayer acts as a mask for the etching process. The nickel layer
was subsequently removed using FeCl$_3$ which does not attack the
gold layer. Finally the combination of diamond and gold layer was
exposed to HF gas for suspension. The etch rate of gold in HF at
the temperatures used is negligible \cite{Williams2003}.

In the following, we mainly discuss the results of two resonators,
one with a geometry of 30 $\mu$m x 480 nm (length x width) and one
with 25 $\mu$m x 350 nm with a 50 nm gold layer on the top,
referred to as sample A and sample B, respectively. The thickness
of the diamond film was estimated to 300 nm using an optical
profilometer.

\section{\label{sec:meas}Measurements}
The low temperature characterization of the nano-mechanical beams
was done using the magneto-motive detection scheme
\cite{Cleland1996}. The radio frequency signal from a network
analyzer (Rohde - Schwarz ZVL-13) was fed into a coaxial line at
the top of a $^3$He cryostat with a base temperature of 500 mK.
The signal was delivered to the sample through two attenuation
stages: 20 dB at 4.2 K and 20 dB at the 1.2 K stage. An ac-current
flowing through the sample exposed to a perpendicular external
magnetic field B induces a Lorentz force that actuates the beam
and leads to a displacement of the nano-mechanical beam in plane
to the diamond film. On resonance, the beam dissipates energy
changing its impedance and resulting in a dip in the transmission
signal. The transmitted signal is amplified at 4.2 K with a gain
of approximately 50 dB (Caltech CITLF1 SN120) and fed into the
input port of the network analyzer. The same electrical set-up
also allows for characterization of the superconducting properties
of the nano-mechanical resonators.

%%%%%%%%%%%%%%%%%%%%%%%%%%%%%%%%%%%%%
%Fig. 2: TRANSMISSION VS. INPUT POWER AND TRANSMISSION VS TEMPERATURE %%%
\begin{figure}[htbp]
\centerline{\includegraphics[width=8.6cm,angle=0]{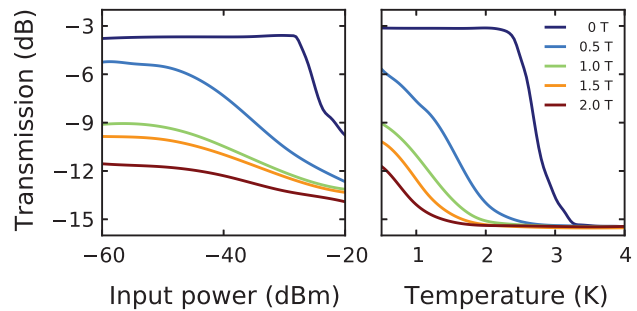}}
\caption{Superconducting characteristics of the resonator. The
diamond resonator (sample B) shows a superconducting state at zero
magnetic field at low input powers (left) and at low temperatures
(right) identified by the plateau region. With increasing input
power (temperature), the sample undergoes a transition into its
normal state, its resistance increases and hence the transmitted
signal decreases. The transition is shifted to lower input powers
(temperatures) at higher magnetic fields.} \label{02}
\end{figure}
%%%%%%%%%%%%%%%%%%%%%%%%%%%%%%%%%%%%%%%%%%%%%%%

To verify whether our resonators show superconductivity, we first
measured the superconducting transition as a function of the input
power and temperature at a frequency of approximately 9 MHz close
to the expected resonance frequency. The input power can in
principle be directly converted into a current using a perfectly
50 Ohm adapted circuit model. However, since this approach
neglects the change of sample impedance when sweeping through the
resonance as well as contact resistances, it is more convenient to
directly plot the input power instead of the bias current.
Nevertheless, the calculated critical currents are of the order of
few $\mu$A, similar to what has been measured with DC measurements
of similar nanostructures made from BDD \cite{MandalN2010,
Mandal2010}. The transmission signal of sample B is plotted in
figure \ref{02}. One can clearly identify a constant transmission
plateau at low input powers (left panel) and at low temperatures
(right panel). A constant transmission directly goes with
unaltered electrical properties for which we can identify the
superconducting state of the beam. The device shows an increase in
impedance at high input powers (temperatures), which leads to a
reduction of the transmitted signal and eventually to the
transition of the sample into its normal state. This impedance
increase is associated with the absorption of microwave power. The
splitting of cooper-pairs leads to the creation of excess
quasiparticles and drastically alters the complex conductivity of
our structure and hence the superconducting state
\cite{deVisser2012}.

From the total transmission drop we can calculate the approximate
normal state resistance values to around 300 Ohms for sample B. We
have obtained similar data for sample A and a normal state
resistance close to 2.5 kOhm (not displayed). The difference in
resistance is due to the presence of the gold layer on top of
sample B. At higher magnetic fields, the superconducting
transition is shifted to lower input powers and to lower
temperatures and a residual resistance appears which can be
associated to the increase of quasiparticles in the
superconductor. We obtained a superconducting transition
temperature of approximately 2.5 K at zero magnetic field in
agreement with measurements on non-suspended diamond samples
\cite{Mandal2011, MandalN2010, Mandal2010}.

\begin{figure}[htbp]
\centerline{\includegraphics[width=8.6cm,angle=0]{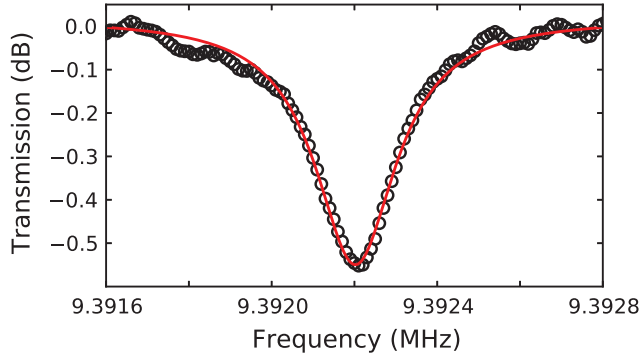}}
\caption{Mechanical resonance of sample A recorded at 2 Tesla
showing a  loaded quality factor of 40\,000. The red line is a
Lorentzian fit.} \label{03}
\end{figure}

We now turn to the mechanical properties of the diamond
nano-mechanical resonator. By applying a perpendicular magnetic
field and sweeping the RF frequency of the bias, the resonators
can be actuated and its characteristics can be extracted. In
figure \ref{03} we show a typical transmission signal at resonance
obtained from sample A. The resonance frequency of resonator A and
B are 9.39226 MHz and 8.77142 MHz respectively. Using
\begin{equation}
f_{res}=\frac{1}{2\pi} \frac{\chi^2}{l^2} \sqrt{\frac{Y I_y}{\rho w t}}
\end{equation}
with $\chi=4.73$ being a numerical factor for the beams' first
flexual mode \cite{cleland2003}, $I_y$ being its moment of
Inertia, w its width, l its length and $\rho$ the density of
diamond, we can calculate the Young's modulus to 950 and 810 GPa,
respectively. The difference between the Young's moduli of sample
A and B is simply due to the additional metal layer of the latter.
In addition, the fact that the Young's modulus is as high as the
one observed for undoped nanocrystalline diamond
\cite{Williams2010, Hees2013} shows that the boron doping does not
degrade the mechanical properties.

%%%%%%%%%%%%%%%%%%%%%%%%%%%%%%%%%%%%%
%Fig. 3: RESONANCE AT 2T, Q = 40\,000, sample I%%%
%%%%%%%%%%%%%%%%%%%%%%%%%%%%%%%%%%%%%

From the transmission measurement we can also extract the loaded
quality factor Q, which describes the rate of energy loss,
compared to the energy stored in the resonator. For sample A and B
we find Q = 40000 and Q = 30000, respectively. From these
measurements we conclude that the Young's modulus as well as the
quality factor of sample B is lower due to enhanced losses caused
by the gold layer on top of the structure. The effect of the gold
layer is to modify the mechanical properties in terms of surface
stress, additional mass, additional elasticity and damping. Such
effects have been studied in detail in the literature
\cite{Collin2010, ImbodenN2013, ImbodenA2013}.
%%%%%%%%%%%%%%%%%%%%%%%%%%%%%%%%%%%%%
%Fig. 4: DAMPING VS FIELD%%%%%%%%%%%%%%%%%%%%%%%%
\begin{figure}[htbp]
\centerline{\includegraphics[width=8.6cm,angle=0]{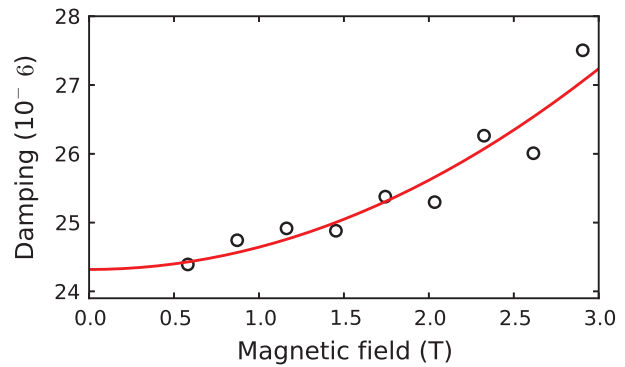}}
\caption{Damping of the mechanical resonator A as a function of
magnetic field.  The quadratic dependence indicates that the
damping is governed by eddy-currents.} \label{04}
\end{figure}
%%%%%%%%%%%%%%%%%%%%%%%%%%%%%%%%%%%%%

A side effect of the magneto-motive detection technique is the
circulation of eddy currents inside the structure, which leads to
an additional magnetic field that is opposed to the applied
external magnetic field. This results in an additive force that is
opposed to the beam movement and leads to another damping term
Q$_E$ that adds linearly to the intrinsic mechanical damping.
\begin{equation}
\frac{1}{Q}=\frac{1}{Q_{mech}}+\frac{1}{Q_{E}}
\end{equation}
The eddy current damping \cite{Huang2005} scales with B$^2$ and
adds to the inverse of the intrinsic quality factor which is
independent of the magnetic field and only depends on the
intrinsic mechanical losses. Figure \ref{04} shows the
corresponding magnetic damping for sample A from which we extract
the intrinsic unloaded quality factor Q$_{mech}$ = 41000 at zero
magnetic field. This mechanical quality factor is limited by the
surface roughness of the diamond and more importantly by clamping
losses due to the doubly-clamped beam design. The isotropic
etching of the sacrificial SiO$_2$ layer leads to an undercut of
the anchor pads of the NEMS. The more surface undercut the more
dissipation is possible in the vibrating surroundings of the
resonators' clamps, the higher the losses. Possible solutions to
increase this quality factor would be to use the so-called
free-free beam design \cite{Huang2005}, to reduce the surface
losses by smoothing the surface with mechanical or chemical
polishing before nanofabrication \cite{Thomas2014} and to remove
the undercut by means of a focused ion beam technique.
Nevertheless, the observed quality factors are comparable with
state-of-the art resonators \cite{Ovartchaiyapong2012, Faust2012,
Schmid2012}. A commonly used value for comparison is the product
of frequency and quality factor, fQ, for which we obtain 3.85
$\times$ 10$^{11}$.

%%%%%%%%%%%%%%%%%%%%%%%%%%%%%%%%%%%%%
%Fig. 5: %SIGNAL AMPLITUDE VS MAGNETIC FIELD SQUARD
\begin{figure}[htbp]
\centerline{\includegraphics[width=8.6cm,angle=0]{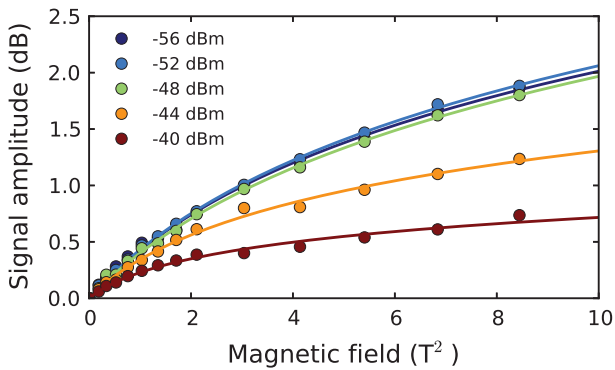}}
\caption{The signal amplitude as a function of magnetic field at
different input powers. A logarithmic scaling with the squared
magnetic field accounts for a mechanical resonance.} \label{05}
\end{figure}
%%%%%%%%%%%%%%%%%%%%%%%%
%%%%%%%%%%%%%%%%%%%%%%%%%%%%%%%%%%%%%
%Fig. 6:: SHIFT IN RESONANCE FREQUENCY AT INCREASING FIELDS
\begin{figure}[htbp]
\centerline{\includegraphics[width=8.6cm,angle=0]{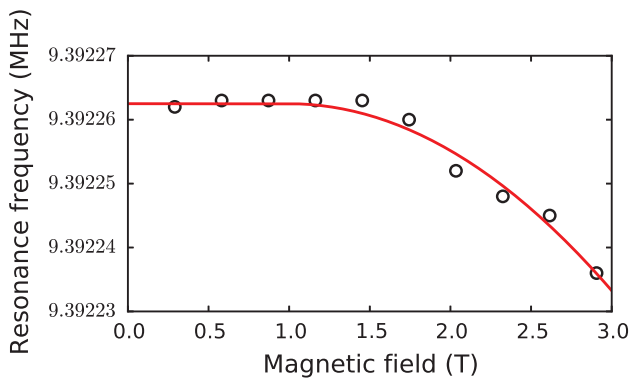}}
\caption{The mechanical center frequency is shifted due to an
embedding impedance that appears around 1T. The data has been
fitted with equation \ref{loadedfreq}.} \label{06}
\end{figure}
%%%%%%%%%%%%%%%%%%%%%%%%%%%%%%%%%%%%%

To convince oneself that the measured resonance is of mechanical
nature and not of some electrical resonance, we plot the signal
amplitude as a function of magnetic field in figure \ref{05}.
Following \cite{Cleland1999}, we can fit our curves using
\begin{equation}
S_{12}=-20 \beta Log[\frac{2 Z_0}{\alpha B^2+2 Z_0}]
\end{equation}
where $\alpha B^2 = Z_c = \frac{\xi l^2 B^2}{\omega_0 m}$ is the
resonators impedance, $\beta$ adjusts the amplitude of the signal,
$Z_0$ is the line impedance and $\xi$ is a constant of order
unity, depending on the mode shape \cite{Cleland1999}.

The transmission signal of the resonator A shows a finite residual
resistance at finite magnetic field as depicted in figure
\ref{02}. This resistance can be seen as the real part of an
external embedding impedance in series with the nano-mechanical
oscillator. Assuming that this embedding impedance changes slowly
over the resonance width which is justified in our case as the
damping is low, we find that the loaded resonance frequency shifts
according to \cite{Cleland1999}
\begin{equation}
f_l = f_0 \sqrt{1+\Theta(B-B_c) Z_c \frac{\Re e(Z_{ext})}{|Z_{ext}^2|}}
\label{loadedfreq}
\end{equation}
where $f_0$ is the unshifted frequency at zero embedding
impedance. Assuming that the second term in equation
\ref{loadedfreq} is zero below a critical field $B_c$ for the
superconducting resonator we can fit our data of the resonance
frequency shift as shown in figure \ref{06}. From the fit we
extract the critical field to $B_c = 0.996 T$ which is consistent
with the onset of the residual resistance of sample A (not
displayed).

\section{\label{sec:concl}Conclusion}
We have demonstrated that nano-mechanical resonators made from
boron doped diamond show superconducting properties up to magnetic
fields of 3 Teslas. These resonators show high quality factors as
high as 40000 at a resonance frequency of around 10 MHz. The
simple fabrication process of superconducting diamond resonators
allows for easy implementation into fully superconducting diamond
circuits such as micro-cavities or superconducting quantum
interference devices. Due to its remarkable mechanical, electrical
as well as optical properties we conclude that nano-mechanical
resonators made from boron doped diamond offer an extremely
attractive system in the growing field of quantum opto-mechanics.
\begin{acknowledgments}
C.B. acknowledges financial support from the French National
Agency (ANR) in the frame of its program in Nanosciences and
Nanotechnologies (SUPERNEMS project no
anr-08-nano-033).\end{acknowledgments}

\bibliography{biblio}

\end{document}